\documentclass[a4paper,10pt,twoside]{article}

\usepackage{amssymb}
\usepackage{amsmath}
\usepackage{amsthm}

\makeatletter
\let\@afterindentfalse\@afterindenttrue
\@afterindenttrue
\makeatother

\newtheorem{theorem}{Theorem}
\newtheorem{lemma}{Lemma}
\newtheorem{corollary}{Corollary}

\newcommand{\la}{\lambda}
\newcommand{\dint}{\mathop{\mathrm{d}}\nolimits\hskip-1pt}
\newcommand{\Ga}{\Gamma}
\newcommand{\E}{\mathop{\mathrm{E}}\nolimits}
\newcommand{\e}{\mathop{\mathrm{e}}\nolimits}
\newcommand{\MR}[1]{\mathcal{M}^{\mathbb{R}}_{#1}}
\newcommand{\MC}[1]{\mathcal{M}^{\mathbb{C}}_{#1}}
\newcommand{\MQ}[1]{\mathcal{M}^{\mathbb{H}}_{#1}}
\newcommand{\MN}{\mathcal{M}_{n}}
\newcommand{\Mt}{\mathcal{M}_{2}}
\newcommand{\Tr}{\mathop{\mathrm{Tr}}\nolimits}
\newcommand{\ci}{\mathop{\mathrm{i}}\nolimits}

\newcommand{\btop}[2]{\genfrac{}{}{0pt}{2}{#1}{#2}}

\title{
  Volume of the quantum mechanical state space\thanks{keywords: state space, volume, monotone metrics;
          MSC: 53C20, 81Q99}}
\author{Attila Andai\thanks{andaia@math.bme.hu}\\
  Department for Mathematical Analysis,\\
  Budapest University of Technology and Economics,\\
  H-1521 Budapest XI. Sztoczek u. 2, Hungary}
\date{April 6, 2006}

\begin{document}

\maketitle

\begin{abstract}
The volume of the quantum mechanical state space over $n$-dimensional real, complex
  and quaternionic Hilbert-spaces with respect to the canonical Euclidean measure is computed,
  and explicit formulas are presented for the expected value of the determinant in the general
  setting too.
The case when the state space is endowed with a monotone metric or a pull-back metric is considered
  too, we give formulas for the volume of the state space with respect to the given
  Riemannian metric.
We present the volume of the space of qubits with respect to various monotone metrics.
It turns out that the volume of the space of qubits can be infinite too.
We characterize those monotone metrics which generates infinite volume.
\end{abstract}

\section*{Introduction}

The classical Jeffreys' prior is the square root of the determinant of the classical Fisher
  information matrix, up to a normalization constant.
Analogously, the quantum mechanical counterpart of the Jeffreys' prior is the square root of
  the determinant of the quantum Fisher-information matrix.
In the quantum mechanical case one can endow the state space with different Riemannian metrics.
Some of them are the monotone metrics, which can be labeled by special operator monotone functions.
In this general case the Jeffreys' prior is still unknown explicitly.
This prior was widely examined by Slater if the metric is the Bures metric
  \cite{Sla1,Sla2,Sla3,Sla4,Sla5,Sla6}.
In this paper we compute the volume of the state space with respect to the Lebesgue-measure,
  we give general formulas for the volume for monotone and pull-back metrics too and
  we characterize those monotone metrics which generate infinite volume of the state of qubits.

In the first section we fix the notations for further computations and we mention some elementary
  lemmas which will be used in the sequel.
In the second section we compute the volume of the quantum mechanical state space over
  $n$-dimensional real, complex and quaternionic Hilbert-spaces with respect to the canonical
  Euclidean measure.
Before these general computations we compute the volume of the state space over the $3$ and $4$
  dimensional real Hilbert-space to give insight into the general computational method.
Moreover, we present explicitly the expected value of the determinant in the general setting.
In the third section we consider the case when the state space is endowed with a monotone metric
  or a pull-back metric and we give formulas to compute the volume of the state space with
  respect to the given Riemannian metric.
Finally, in the fourth section we deal with the qubit case.
We present the volume of this space with respect to various monotone metrics.
It turns out that the volume of the space of qubits can be infinite too.
We characterize those monotone metrics which generate infinite volume.

\section{Basic lemmas and notations}

The quantum mechanical state space consists of real, complex or quaternionic self-adjoint positive
  matrices with trace $1$.
We consider only the set of faithful states with real, complex and quaternionic entries.
\begin{align*}
&\MR{n}=\left\{X\in M(n,\mathbb{R}) \ \vert\ X=X^{*}, X>0, \Tr X=1 \right\}\\
&\MC{n}=\left\{X\in M(n,\mathbb{C}) \ \vert\ X=X^{*}, X>0, \Tr X=1 \right\}\\
&\MQ{n}=\left\{X\in M(n,\mathbb{H}) \ \vert\ X=X^{*}, X>0, \Tr X=1 \right\}
\end{align*}

The following lemmas will be our main tools, we will use them without mentioning,
  and we also introduce some notations which will be used in the sequel.

The first lemma is about some elementary properties of the gamma function $\Gamma$.
\begin{lemma}
Consider the function $\Ga$, which can be defined for $z\in\mathbb{R}^{+}$ as
\begin{equation*}
\Ga(z)=\int_{0}^{\infty}t^{z-1}\e^{-t}\dint t.
\end{equation*}
This function has the following properties for every natural number $n\neq 0$ and
  real argument $z\in\mathbb{R}^{+}$.
\begin{align*}
&\Ga(n)=(n-1)!\quad \Ga(1+z)=z\Ga(z)\quad \Ga(1/2)=\sqrt{\pi}\\
&\Ga(n+1/2)=\frac{(2n-1)!!}{2^{n}}\sqrt{\pi}\quad \Ga(n/2)=\frac{(n-2)!!}{2^{\frac{n-1}{2}}}\sqrt{\pi}
\end{align*}
\end{lemma}

For an $n\times n$ matrix $A$ we set $A_{i}$ to be the left upper $i\times i$ submatrix of $A$,
  where $i=1,\dots,n$.
The next two lemmas are elementary proposition in linear algebra.

\begin{lemma}
The $n\times n$ self-adjoint matrix $A$ is positive definite if and only if the inequality
  $\det(A_{i})>0$ holds for every $i=1,\dots,n$.
\end{lemma}

\begin{lemma}
Assume that $A$ is an $n\times n$ matrix with entries $x$ ($x\in\mathbb{R}$) and $B$ is a diagonal matrix
  with the elements $B_{jj}$ on the main diagonal, then
\begin{equation*}
\det(A+B)=\det(B)+x\sum_{i=1}^{n}\prod_{\btop{j=1}{j\neq i} }^{n}B_{jj}.
\end{equation*}
\end{lemma}

\begin{lemma}\label{le:determinant}
Assume that $A$ is an $n\times n$ self-adjoint, positive definite matrix with entries $(a_{ij})_{i,j=1,\dots, n}$
  and the vector $x$ consists of the first $(n-1)$ elements of the last column, that is $x=(a_{1,n},\dots,a_{n-1,n})$.
Then for the matrix $T=\det(A_{n-1}) (A_{n-1})^{-1}$ we have
\begin{equation*}
\det(A)=a_{nn}\det(A_{n-1})-\left< x,Tx \right>.
\end{equation*}
\end{lemma}
\begin{proof}
Elementary matrix computation, one should expand $\det(A)$ by minors, with respect to the last row.
\end{proof}

\begin{lemma}
For parameters $a,b\in\mathbb{R}^{+}$ and $t\in\mathbb{R}^{+}$ the integral equalities
\begin{align*}
         &\int_{0}^{t} x^{a}(t-x)^{b}\dint x=t^{1+a+b}\frac{\Ga(a+1)\Ga(b+1)}{\Ga(a+b+2)}\\
G_{a,b}:=&\int_{0}^{1} x^{a}(1-x^{2})^{b}\dint x
  =\frac{1}{2}\frac{\Ga(b+1)\Ga\left(\frac{a+1}{2}\right)}{\Ga\left(\frac{a}{2}+b+\frac{3}{2}\right)}
\end{align*}
hold.
\end{lemma}
\begin{proof}
These are consequences of the formula below for the beta integral
\begin{equation*}
\int_{0}^{1}x^{p}(1-x)^{q}\dint x=\frac{\Ga(p+1)\Ga(q+1)}{\Ga(p+q+2)}.
\end{equation*}
\end{proof}

\begin{lemma}
The surface $F_{n-1}$ of a unit sphere in an $n$ dimensional space is
\begin{equation*}
F_{n-1}=\frac{n\pi^{\frac{n}{2}}}{\Ga\left(\frac{n}{2}+1 \right)}.
\end{equation*}
\end{lemma}
\begin{proof}
It follows from the well-known formula for the volume of the sphere in $n$ dimension with radius $r$
\begin{equation*}
V_{n}(r)=\frac{r^{n}\pi^{\frac{n}{2}}}{\Ga\left(\frac{n}{2}+1 \right)},
\end{equation*}
since $F_{n-1}=\left.\frac{\dint V_{n}(r)}{\dint r}\right\vert_{r=1}$.
\end{proof}

When we integrate on a subset of the Euclidean space we always integrate with respect to the usual
  Lebesgue measure.
The Lebesgue measure on $\mathbb{R}^{n}$ will be denoted by $\la_{n}$.

\begin{lemma}
Consider the simplex
\begin{equation*}
\Delta_{n-1}=\left\{ (x_{1},\dots,x_{n})\in\left]0,1\right[^{n}\ \left\vert\ \sum_{k=1}^{n}x_{k}=1 \right.\right\},
\end{equation*}
then
\begin{equation*}
\int\limits_{\Delta_{n-1}} \left(\prod_{i=1}^{n}x_{i}\right)^{k}\dint\la_{n-1}(x)=\frac{\Ga(k+1)^{n}}{\Ga(n(k+1))}.
\end{equation*}
\end{lemma}
\begin{proof}
The integral can be computed as
\begin{align*}
&\int\limits_{\Delta_{n-1}} \left(\prod_{i=1}^{n}x_{i}\right)^{k}\dint\la_{n-1}(x)\\
&=\int\limits_{0}^{1}\int\limits_{0}^{1-a_{1}}\dots\hskip-1em\int\limits_{0}^{1-\sum_{j=1}^{n-2}a_{j}}\hskip-0.5em
  \left(\prod_{i=1}^{n-1}a_{i}^{k}\right)
  \left[ \left(1-\sum_{i=1}^{n-2}a_{i}\right)-a_{n-1} \right]^{k}
  \dint a_{n-1}\dots \dint a_{2}\dint a_{1}.
\end{align*}
Integrating with respect to $a_{n-1}$, the integral is
\begin{equation*}
\frac{\Ga(k+1)\Ga(k+1)}{\Ga(2k+2)}a_{1}^{k}a_{2}^{k}\dots a_{n-2}^{k}
  \bigl((1-a_{1}-\dots-a_{n-3})-a_{n-2} \bigr)^{2k+1}
\end{equation*}
and in general, the $i$-th integral is
\begin{equation*}
\frac{\Ga(k+1)\Ga(ik+i)}{\Ga((i+1)(k+1))}a_{1}^{k}a_{2}^{k}\dots a_{n-1-i}^{k}
  \bigl((1-a_{1}-\dots-a_{n-2-i})-a_{n-1-i} \bigr)^{(i+1)k+i}.
\end{equation*}
Thus the result is
\begin{align*}
&\frac{\Ga(k+1)\Ga(k+1)}{\Ga(2k+2)} \frac{\Ga(k+1)\Ga(2k+2)}{\Ga(3k+3)}
  \frac{\Ga(k+1)\Ga(3k+3)}{\Ga(4k+4)}\times\dots\\
&\qquad  \times\frac{\Ga(k+1)\Ga((n-1)(k+1))}{\Ga(n(k+1))}=\frac{\Ga(k+1)^{n}}{\Ga(n(k+1))}.
\end{align*}
\end{proof}

\begin{lemma}
Assume that $T$ is an $n\times n$ self-adjoint, positive definite matrix and $k,\rho\in\mathbb{R}^{+}$.
Set
\begin{align*}
&\E^{\mathbb{R}}_{n}(T,\rho)=\left\{x\in\mathbb{R}^{n}\ \vert\ \left<x,Tx\right><\rho \right\},
  \quad T_{ij}\in \mathbb{R};\\
&\E^{\mathbb{C}}_{n}(T,\rho)=\left\{x\in\mathbb{C}^{n}\ \vert\ \left<x,Tx\right><\rho \right\},
  \quad T_{ij}\in \mathbb{C};\\
&\E^{\mathbb{H}}_{n}(T,\rho)=\left\{x\in\mathbb{H}^{n}\ \vert\ \left<x,Tx\right><\rho \right\},
  \quad T_{ij}\in \mathbb{H};
\end{align*}
then
\begin{align*}
&\int\limits_{E^{\mathbb{R}}_{n}(T,\rho)}(\rho-\left< x,Tx\right>)^{k}\dint\la_{n}(x)
  =\frac{\rho^{\frac{n}{2}+k}}{\sqrt{\det(T)}}F_{n-1}G_{n-1,k},\\
&\int\limits_{E^{\mathbb{C}}_{2n}(T,\rho)}(\rho-\left< x,Tx\right>)^{k}\dint\la_{2n}(x)
  =\frac{\rho^{n+k}}{\det(T)}F_{2n-1}G_{2n-1,k},\\
&\int\limits_{E^{\mathbb{H}}_{4n}(T,\rho)}(\rho-\left< x,Tx\right>)^{k}\dint\la_{4n}(x)
  =\frac{\rho^{2n+k}}{\det(T)^{2}}F_{4n-1}G_{4n-1,k}.
\end{align*}
\end{lemma}
\begin{proof}
We prove the statement for the real case only, the other cases can be proved in the same way.
The set $\E^{\mathbb{R}}_{n}(T,\rho)$ is a $n$ dimensional ellipsoid, so to compute the integral first we transform
  our canonical basis to a new one, which is parallel to the axes of the ellipsoid.
Since this is an orthogonal transformation, it's Jacobian is $1$.
When we transform this ellipsis to a unit sphere, the Jacobian of this transformation is
\begin{equation*}
\prod_{k=1}^{n}\sqrt{\frac{\rho}{\mu_{k}}},
\end{equation*}
  where $(\mu_{k})_{k=1,\dots,n}$ are the eigenvalues of $T$.
Then we compute the integral in spherical coordinates.
The integral with respect to the angles give the surface of the sphere $F_{n-1}r^{n-1}$ and the radial part is
\begin{equation*}
\int_{0}^{1}F_{n-1}r^{n-1}\sqrt{\frac{\rho^{n}}{\det(T)}}(\rho-\rho r^{2})^{k} \dint r.
\end{equation*}
\end{proof}

\section{Volume of the state space with respect to the Lebesgue measure}

Before investigating the general setting we compute the volume of the spaces $\MR{3}$ and $\MR{4}$.
For a matrix with real entries
\begin{equation*}
A=\begin{pmatrix} a & f & h\\ f & b & g \\ h & g & c\end{pmatrix}
\end{equation*}
we set $A_{1}=(a)$, $A_{2}=\begin{pmatrix} a & f \\ f & b\end{pmatrix}$, $A_{3}=A$ and $D$ denotes
  the matrix, which contains only the diagonal elements of $A$, that is $D_{ij}=\delta_{ij}A_{ii}$.
The matrix $A$ is in $\MR{3}$ if and only if the following set of inequalities hold
\begin{align*}
&\det(A_{1})=a>0 \quad a+b+c=1\\
&\det(A_{2})=ab-f^{2}> 0\\
&\det(A_{3})=abc+2fgh-h^{2}b-g^{2}a-f^{2}c> 0.
\end{align*}
These inequalities can be rewritten as
\begin{equation*}
(a,b,c)\in\Delta_{2},\quad \left< (f),T_{1}(f)\right><b\det(A_{1}),
  \quad\left< (h,g),T_{2}(h,g)\right><c\det(A_{2}),
\end{equation*}
  where $T_{i}=\det(A_{i})A_{i}^{-1}$ for $i=1,2$.
It means that for fixed $D$ and $A_{2}$ the parameters $(h,g)$ are in $\E^{\mathbb{R}}_{2}(T_{2},c\det(A_{2}))$.
We set $V(A_{2})$ to be equal of the volume of the parameter space of $(h,g)$ if $D$ and $A_{2}$ are given, that is
\begin{equation*}
V(A_{2})=\hskip-1em\int\limits_{\E^{\mathbb{R}}_{2}(T_{2},c\det(A_{2}))}\hskip-1em 1 \dint g\dint h
  =\frac{c\det(A_{2})}{\sqrt{\det(T_{2})}}\frac{\pi}{\Ga(2)}=\pi c\sqrt{\det(A_{2})}
\end{equation*}
If $D$ and $A_{1}$ are fixed then we set $V(A_{1})$ to be equal of the volume of the parameter space $(f,g,h)$:
\begin{equation*}
V(A_{1})=\int\limits_{\E^{\mathbb{R}}_{1}(T_{1},b\det(A_{1}))} V(A_{2}) \dint f
  =\int\limits_{-\sqrt{ab}}^{\sqrt{ab}}\pi c\sqrt{ab-f^{2}}\dint f=\frac{\pi^{2}}{2}abc.
\end{equation*}
Finally the volume of the $\MR{3}$ space is
\begin{equation*}
V(\MR{3})=\int\limits_{\Delta_{2}}V(A_{1})\dint\la_{2}=\frac{\pi^2}{2}\int_{0}^{1}\int_{0}^{1-a}ab(1-a-b)\dint b\dint a
=\frac{\pi^{2}}{240}.
\end{equation*}

A $4\times 4$ real, symmetric matrix with diagonal elements $a_{1},a_{2},a_{3}$ and $a_{4}$ b
  is an element of the space $\MR{4}$ if and only if
\begin{align*}
&\det(A_{1})=a_{1}> 0\quad \sum_{k=1}^{4}a_{k}=1\\
&\det(A_{2})=a_{2}\det(A_{1})-\left<x_{1},T_{1}x_{1}\right>>0\\
&\det(A_{3})=a_{3}\det(A_{2})-\left<x_{2},T_{2}x_{2}\right>>0\\
&\det(A_{4})=a_{4}\det(A_{3})-\left<x_{3},T_{3}x_{3}\right>>0.
\end{align*}
We set $T_{i}=\det(A_{i})A_{i}^{-1}$ for $i=1,2,3$.
For fixed parameters $A_{3}$ and $D$
\begin{equation*}
V(A_{3})=\hskip-1em\int\limits_{\E^{\mathbb{R}}_{3}(T_{3},a_{4}\det(A_{3}))}\hskip-1em 1\dint \la_{3}=a_{4}^{3/2}F_{2}G_{2,0}\sqrt{\det(A_{3})},
\end{equation*}
where we used the notation of the previous example, in this case $V(A_{3})$ is the volume the space of those parameters
  which do not belong to $A_{3}$ and $D$.
Now assume that $A_{2}$ and $D$ is given, then
\begin{align*}
V(A_{2})&=\int\limits_{\E^{\mathbb{R}}_{2}(T_{2},a_{3}\det(A_{2}))}  a_{4}^{3/2}F_{2}G_{2,0}\sqrt{\det(A_{3})} \dint \la_{2}\\
&=F_{2}G_{2,0}a_{4}^{3/2}\hskip-1em\int\limits_{\E^{\mathbb{R}}_{2}(T_{2},a_{3}\det(A_{2}))}\hskip-1em
  \bigl(a_3\det(A_{2})-\left<x,T_{2}x\right>\bigr)^{\frac{1}{2}}\dint\la_{2}(x)\\
&=F_{2}F_{1}G_{2,0}G_{1,1/2}a_{4}^{3/2}a_{3}^{3/2}\det(A_{2}).
\end{align*}
If $D$ is given
\begin{align*}
&V(A_{1})=\hskip-1em\int\limits_{\E^{\mathbb{R}}_{1}(T_{1},a_{2}\det(A_{1}))}\hskip-1em
   F_{2}F_{1}G_{2,0}G_{1,1/2}a_{4}^{3/2}a_{3}^{3/2}\det(A_{2})    \dint \la_{1}\\
&=F_{2}F_{1}G_{2,0}G_{1,1/2}a_{4}^{3/2}a_{3}^{3/2}\hskip-1em\int\limits_{\E^{\mathbb{R}}_{1}(T_{1},a_{2}\det(A_{1}))}\hskip-1em
  \bigl(a_2\det(A_{1})-\left<x,T_{1}x\right>\bigr)\dint\la_{1}(x)\\
&=F_{2}F_{1}F_{0}G_{2,0}G_{1,1/2}G_{0,1}(a_{1}a_{2}a_{3}a_{4})^{3/2}.
\end{align*}
Since
\begin{equation*}
\int\limits_{\Delta_{3}} (a_{1}a_{2}a_{3}a_{4})^{3/2}\dint\la_{3}(a)=\frac{\Ga(3/2+1)^{4}}{\Ga(10)}
\end{equation*}
the volume of the $4$ dimensional real state space is
\begin{equation*}
V(\MR{4})=F_{2}F_{1}F_{0}G_{2,0}G_{1,1/2}G_{0,1}\frac{\Ga(3/2+1)^{4}}{\Ga(10)}=\frac{3\pi^{4}}{8\cdot 9!}.
\end{equation*}

\begin{theorem}
For every $k\in\mathbb{N}$ the volume of the state spaces $\MR{2k}$ and $\MR{2k+1}$ are
\begin{align*}
&V(\MR{2k})=\frac{\pi^{k^{2}}}{2^{k^{2}+k}}\frac{(2k)!}{k!(2k^{2}+k-1)!}\prod_{i=1}^{k-1}(2i)! \\
&V(\MR{2k+1})=\left(\frac{\pi}{2}\right)^{k^{2}+k}\frac{(2k)!}{(k-1)!(2k^{2}+3k)!}\prod_{i=1}^{k-1}(2i)!\ .
\end{align*}
\end{theorem}
\begin{proof}
A self-adjoint $n\times n$ matrix with real entries $A$ is in $\MR{n}$ if and only if
\begin{equation*}
\forall i\in\left\{1,\dots,n\right\}:\quad \det(A_{i})>0,\quad \sum_{k=1}^{n}a_{k}=1,
\end{equation*}
  where $(a_{i})_{i=1,\dots,n}$ are the diagonal elements of $A$.
First we assume that the matrix of the diagonal elements, $D$ is given.
If $A_{n-1}$ is fixed, then
\begin{equation*}
V(A_{n-1})=\hskip-1em\int\limits_{\E^{\mathbb{R}}_{n-1}(T_{n-1},a_{n}\det(A_{n-1}))}\hskip-3em 1\dint\la_{n-1}
  =a_{n}^{(n-1)/2}F_{n-2}G_{n-2,0}\sqrt{\det(A_{n-1})}.
\end{equation*}
If $A_{n-2}$ is fixed, then
\begin{align*}
V(A_{n-2})&=\hskip-4em\int\limits_{\E^{\mathbb{R}}_{n-2}(T_{n-2},a_{n-1}\det(A_{n-2}))}\hskip-4em V(A_{n-1})\dint\la_{n-2}\\
&=a_{n}^{\frac{n-1}{2}}F_{n-2}G_{n-2,0}\hskip-4em\int\limits_{\E^{\mathbb{R}}_{n-2}(T_{n-2},a_{n-1}\det(A_{n-2}))}\hskip-4em
    \bigl(a_{n-1}\det(A_{n-2})-\left<x,T_{n-2}x\right>\bigr)^{\frac{1}{2}}\dint\la_{n-2}(x)\\
&\quad=a_{n-1}^{\frac{n-1}{2}}a_{n}^{\frac{n-1}{2}}F_{n-2}F_{n-3}G_{n-2,0}G_{n-3,1/2}\det(A_{n-2}).
\end{align*}
In general if $A_{n-k}$ is fixed, then
\begin{equation*}
V(A_{n-k})=\prod_{i=1}^{k}\Bigl(a_{n+1-i}^{(n-1)/2} F_{n-1-i}G_{n-1-i,(i-1)/2} \Bigr)\det(A_{n-k})^{\frac{k}{2}},
\end{equation*}
because this equation is correct for $k=1$ and by induction
\begin{align*}
&\hskip-4em\int\limits_{\E^{\mathbb{R}}_{n-k-1}(T_{n-k-1},a_{n-k}\det(A_{n-k-1}))}\hskip-4em V(A_{n-k})\dint\la_{n-k-1}\\
&=\prod_{i=1}^{k}\Bigl(a_{n+1-i}^{(n-1)/2} F_{n-1-i}G_{n-1-i,(i-1)/2} \Bigr)\\
&\quad\times  \hskip-4em\int\limits_{\E^{\mathbb{R}}_{n-k-1}(T_{n-k-1},a_{n-k}\det(A_{n-k-1}))}\hskip-4em
  \bigl(a_{n-k}\det(A_{n-k-1})-\left<x,T_{n-k-1}x\right>\bigr)^{\frac{k}{2}}\dint\la_{n-k-1}(x)\\
&=\prod_{i=1}^{k}\Bigl(a_{n+1-i}^{(n-1)/2} F_{n-1-i}G_{n-1-i,(i-1)/2} \Bigr)\\
&\quad \times a_{n-k}^{(n-1)/2}F_{n-k-2}G_{n-k-2,k/2}\det(A_{n-k-1})^{\frac{k+1}{2}}=V(A_{n-k-1}).
\end{align*}
It means that
\begin{equation*}
V(A_{1})=\left(\prod_{i=0}^{n-2}F_{i}\right) \left(\prod_{i=1}^{n-1}G_{n-1-i,(i-1)/2} \right)
 \left(\prod_{i=1}^{n}a_{n}\right)^{(n-1)/2}.
\end{equation*}
So the volume of the real state space is
\begin{equation*}
V(\MR{n})=\left(\prod_{i=1}^{n-1}F_{i-1}G_{n-1-i,(i-1)/2} \right)
 \int\limits_{\Delta_{n-1}}\left(\prod_{i=1}^{n}a_{n}\right)^{(n-1)/2}\dint\la_{n-1}(a).
\end{equation*}
The integral in this equation is
\begin{equation*}
\frac{\Ga\left(\frac{n+1}{2}\right)^{n}}{\Ga\left(\frac{n^{2}+n}{2} \right)}
\end{equation*}
and the product is
\begin{equation*}
\varphi=\left(\prod_{i=1}^{n-1}F_{i-1}G_{n-1-i,(i-1)/2} \right)
=\frac{\pi^{\frac{n^{2}-n}{4}}}{2^{n-1}}\frac{(n-1)!}{\Ga\left(\frac{n+1}{2} \right)^{n-1}}
\prod_{i=1}^{n-1}\frac{\Ga\left(\frac{i+1}{2}\right)\Ga\left(\frac{n-i}{2}\right)}{\Ga\left(\frac{i}{2}+1\right)}.
\end{equation*}
If $n=2k+1$, then
\begin{equation*}
\varphi=\frac{\pi^{k^2+\frac{k}{2}}}{2^{2k}}\frac{(2k)!}{(k!)^{2k+1}}\prod_{i=1}^{2k}\Ga\left(\frac{i}{2}\right)
\end{equation*}
which can be simplified using the equality $\Ga(i)\Ga(i+1/2)=\dfrac{\sqrt{\pi}(2i)!}{2^{2i}i}$ to
\begin{equation*}
\varphi=\left(\frac{\pi}{2}\right)^{k^2+k}\frac{k(2k)!}{(k!)^{2k+2}}\prod_{i=1}^{k-1}(2i)!\ .
\end{equation*}
If $n=2k$, then using the same identity for the function $\Ga$ we have
\begin{equation*}
\varphi=\pi^{k^2-k}2^{3k^{2}-k}\left(\frac{k!}{(2k)!}\right)^{2k-1}  \prod_{i=1}^{k-1}(2i)!\ .
\end{equation*}
\end{proof}

\begin{theorem}
For every $n\in\mathbb{N}$ the volume of the state space $\MC{n}$ is
\begin{equation*}
V(\MC{n})=\frac{\pi^{\frac{n(n-1)}{2}}}{(n^{2}-1)!}\prod_{i=1}^{n-1}i!\ .
\end{equation*}
\end{theorem}
\begin{proof}
The proof is similar to the real case, except that we have to take into account that the
  dimension of the parameter space of a matrix element $A_{ij}$ for $i\neq j$
  indices is $2$.
If $A_{n-1}$ is fixed, then
\begin{equation*}
V(A_{n-1})=\hskip-1em\int\limits_{\E^{\mathbb{C}}_{2n-2}(T_{n-1},a_{n}\det(A_{n-1}))}\hskip-3em 1\dint\la_{2n-2}
  =a_{n}^{n-1}F_{2n-3}G_{2n-3,0}\det(A_{n-1})
\end{equation*}
and in general if $A_{n-k}$ is fixed, then
\begin{equation*}
V(A_{n-k})=\prod_{i=1}^{k}\Bigl(a_{n+1-i}^{n-1} F_{2n-1-2i}G_{2n-1-2i,i-1} \Bigr)\det(A_{n-k})^{k}.
\end{equation*}
The volume of the complex state space is
\begin{equation*}
V(\MC{n})=\left(\prod_{i=1}^{n-1}F_{2i-1}G_{2n-1-2i,i-1} \right)
 \int\limits_{\Delta_{n-1}}\left(\prod_{i=1}^{n}a_{n}\right)^{n-1}\dint\la_{n-1}(a),
\end{equation*}
where the product is
\begin{equation*}
\frac{\pi^{\frac{n^{2}-n}{2}}}{((n-1)!)^{n}}\prod_{i=1}^{n-1}i!
\end{equation*}
and the integral is
\begin{equation*}
\frac{((n-1)!)^{n}}{(n^{2}-1)!}.
\end{equation*}
\end{proof}

\begin{theorem}
For every $n\in\mathbb{N}$ the volume of the state space $\MQ{n}$ is
\begin{equation*}
V(\MQ{n})=\frac{(2n-2)!\pi^{n^{2}-n}}{(2n^{2}-n-1)!}\prod_{i=1}^{n-2}(2i)!.
\end{equation*}
\end{theorem}
\begin{proof}
If $A_{n-1}$ is fixed, then
\begin{equation*}
V(A_{n-1})=a_{n}^{2n-2}F_{4n-5}G_{4n-5,0}\det(A_{n-1})^{2}
\end{equation*}
and in general if $A_{n-k}$ is fixed, then
\begin{equation*}
V(A_{n-k})=\prod_{i=1}^{k}\Bigl(a_{n+1-i}^{2n-2} F_{4n-1-4i}G_{4n-1-4i,2i-2} \Bigr)\det(A_{n-k})^{2k}.
\end{equation*}
The volume of the quaternionic state space is
\begin{equation*}
V(\MQ{n})=\left(\prod_{i=1}^{n-1}F_{4i-1}G_{4n-1-4i,2i-2} \right)
 \int\limits_{\Delta_{n-1}}\left(\prod_{i=1}^{n}a_{n}\right)^{2n-2}\dint\la_{n-1}(a),
\end{equation*}
where the product is
\begin{equation*}
\frac{\pi^{n^{2}-n}}{((2n-2)!)^{n-1}}\prod_{i=1}^{n-1}(2i-2)!
\end{equation*}
and the integral is
\begin{equation*}
\frac{((2n-2)!)^{n}}{(2n^{2}-n-1)!}.
\end{equation*}
\end{proof}

A slight modification of the previous proofs gives the following Theorem.

\begin{theorem}
For every parameter $\alpha\in\mathbb{R}^{+}$ and $n\in\mathbb{N}$ the expected value of the
function $\det^{\alpha}$ on the state spaces $\MR{n}$, $\MC{n}$ and $\MQ{n}$ with respect to the normalized
Lebesgue measures $\mu_{\mathbb{R}}$, $\mu_{\mathbb{C}}$ and $\mu_{\mathbb{H}}$  are
\begin{align*}
&\int\limits_{\MR{n}}\det(A)^{\alpha}\dint\mu_{\mathbb{R}}(A)
  =\frac{\Ga\left(\frac{n^{2}+n}{2}\right)}{\Ga\left(\frac{n+1}{2}\right)}
  \frac{\Ga\left(\frac{n+1}{2}+\alpha\right)}{\Ga\left(\frac{n^{2}+n}{2}+n\alpha \right)}
  \prod_{i=1}^{n-1}\frac{\Ga\left(\frac{i+1}{2}+\alpha\right)}{\Ga\left(\frac{i+1}{2}\right)}\\
&\int\limits_{\MC{n}}\det(A)^{\alpha}\dint\mu_{\mathbb{C}}(A)
  =\frac{(n^{2}-1)!}{(n-1)!}\frac{\Ga(n+\alpha)}{\Ga(n^{2}+n\alpha)}
  \prod_{i=1}^{n-1}\frac{\Ga(i+\alpha)}{\Ga(i)}\\
&\int\limits_{\MQ{n}}\det(A)^{\alpha}\dint\mu_{\mathbb{H}}(A)
  =\frac{\Ga(2n^{2}-n)}{\Ga(2n-1)}\frac{\Ga(2n+\alpha-1)}{\Ga(2n^{2}-n+\alpha n)}
  \prod_{i=1}^{n-1}\frac{\Ga(2i-1+\alpha)}{\Ga(2i-1)}.
\end{align*}
\end{theorem}
\begin{proof}
The proofs are similar, so we just prove the theorem for the real case only.
First we compute the integral with respect to the Lebesgue measure, and we divide the result with
  the volume of the state space.
The method is the same as in the previous theorems, so if $A_{n-1}$  is given, then
\begin{align*}
V(A_{n-1})&=\hskip-1em\int\limits_{\E^{\mathbb{R}}_{n-1}(T_{n-1},a_{n}\det(A_{n-1}))}\hskip-3em
  \det(A_{n})^{\alpha}\dint\la_{n-1}\\
&=\hskip-1em\int\limits_{\E^{\mathbb{R}}_{n-1}(T_{n-1},a_{n}\det(A_{n-1}))}\hskip-3em
 \bigl(a_{n}\det(A_{n-1})-\left<x,T_{n-1}x\right>\bigr)^{\alpha}\dint\la_{n-1}(x)\\
  &=a_{n}^{(n-1)/2+\alpha}F_{n-2}G_{n-2,\alpha}\det(A_{n-1})^{\frac{1}{2}+\alpha}
\end{align*}
and the general formula is
\begin{equation*}
V(A_{n-k})=\prod_{i=1}^{k}\Bigl(a_{n+1-i}^{(n-1)/2+\alpha}F_{n-1-i}G_{n-1-i,(i-1)/2+\alpha} \Bigr)
  \det(A_{n-k})^{\frac{k}{2}+\alpha}.
\end{equation*}
The integral of the function $\det^{\alpha}$ with respect to the Lebesgue measure is
\begin{equation*}
\int\limits_{\MR{n}}\hskip-0.5em\det(A)^{\alpha}\dint\la_{\dim(\MR{n})}(A)=\hskip-0.5em
\left(\prod_{i=1}^{n-1}F_{i-1}G_{n-1-i,(i-1)/2+\alpha} \right)
  \hskip-0.5em\int\limits_{\Delta_{n-1}}\hskip-1em\left(\prod_{i=1}^{n}a_{n}\right)^{\frac{n-1}{2}+\alpha}
  \hskip-2.5em\dint\la_{n-1}(a).
\end{equation*}
Dividing it with $V(\MR{n})$ and after some simplification we get the formula which is in the Theorem.
\end{proof}

\section{Volume of the state space endowed with Riemannian metrics}

To simplify the notations the set of real or complex self-adjoint matrices will be denoted by $M_{n}$,
  and the set of real or complex states by $\MN$.

\v{C}encov and Morozova \cite{Cen,CenMor} were the first to study the monotone metrics on classical
  statistical manifolds, they proved that such a metric is unique, up to normalization.
The noncommutative extension of the \v{C}encov Theorem was given by Petz \cite{Pet1}.
Stochastic maps are the counterpart of Markovian maps in this setting.
A linear map between matrix spaces $T:M_{n}\to M_{m}$ is called a stochastic map if it is
  trace preserving and completely positive.

\begin{theorem}
Consider the family of Riemannian-manifolds $(\MN,g_{n})_{n\in\mathbb{N}}$.
If for every stochastic map $T:M_{n}\to M_{m}$ the following monotonicity property holds
\[K_{T(D)}(T(X),T(X))\leq K_{D}(X,X)
  \qquad \forall D,X\in M_{n}\]
  then there exists an operator monotone function $f:\mathbb{R}^{+}\to\mathbb{R}$
  with the property $f(x)=xf(x^{-1})$, such that
\begin{equation*}
g_{D}(X,Y)=\Tr\biggl( X\bigl(
R^{\frac{1}{2}}_{n,D}f(L_{n,D}R^{-1}_{n,D})R^{\frac{1}{2}}_{n,D}\bigr)^{-1}
(Y) \biggr)\ ,
\end{equation*}
  for all $n\in\mathbb{N}$ where $L_{n,D}(X)=DX$, $R_{n,D}(X)=XD$ for all $D,X\in M_{n}$.
\end{theorem}

These metrics are considered as the noncommutative generalizations of the Fisher-information.
These metrics are called monotone metrics.
It means that there exists a bijective mapping between the monotone family of metrics
  and some operator monotone functions.
We use the normalization condition $f(1)=1$ for the function $f$ in the previous theorem.

Let $D\in\MN$ and choose a basis of $\mathbb{R}^{n}$ such that
  $\displaystyle D=\sum_{j=1}^{n}\mu_{j}E_{jj}$ is diagonal, where $(E_{jk})_{1\leq j,k\leq n}$ is the usual
  system of matrix units.
Let us define the following self-adjoint matrices.
\begin{align*}
&F_{jk}=E_{jk}+E_{kj}          &&1\leq j\leq k\leq n\\
&H_{jk}=\ci E_{jk}-\ci E_{kj}  &&1\leq j< k\leq n
\end{align*}
The set of matrices $(F_{ij})_{1\leq i\leq j\leq n}\cup (H_{ij})_{1\leq i<j\leq n}$ form a basis of the
  tangent space at $D$ for complex matrices and $(F_{ij})_{1\leq i\leq j\leq n}$ form a basis for real ones.
We have for the metric from \cite{MPA} that
\begin{align*}
&\mbox{if} &&1\leq i<j\leq n, 1\leq k<l\leq n: \quad
  &&\left\lbrace\begin{array}{l} g(D)(H_{ij},H_{kl})=\delta_{ik}\delta_{jl}2m(\mu_{i},\mu_{j})\\
                               g(D)(F_{ij},F_{kl})=\delta_{ik}\delta_{jl}2m(\mu_{i},\mu_{j})\\
                               g(D)(H_{ij},F_{kl})=0,\end{array}\right.\\
&\mbox{if} &&1\leq i<j\leq n, 1\leq k\leq n: \quad &&g(D)(H_{ij},F_{kk})=g(D)(F_{ij},F_{kk})=0,\\
&\mbox{if} &&1\leq i\leq n,   1\leq k\leq n: \quad &&g(D)(F_{ii},F_{kk})=\delta_{ik}4m(\mu_{i},\mu_{i}),
\end{align*}
where
\begin{equation*}
m(\mu_{i},\mu_{j})=\frac{1}{\mu_{j}f\left(\dfrac{\mu_{i}}{\mu_{j}} \right)}.
\end{equation*}

The volume of a Riemannian manifold $(M,g)$ is defined as
\begin{equation*}
\int_{M}\sqrt{\det(g(x))}\dint\la_{\dim(M)}(x).
\end{equation*}

We use the canonical parametrization for the off-diagonal elements of $\MN$ and
  $(x_{1},\dots,x_{n-1},1-(x_{1}+\dots+x_{n-1}))$ for the diagonal ones.
The corresponding tangent vectors for the diagonal coordinates are $A_{i}=E_{ii}-E_{nn}$ for $i=1,\dots,n-1$.
Since $g(D)(A_{i},A_{j})=g(D)(E_{ii},E_{jj})+g(D)(E_{nn},E_{nn})=\delta_{ij}\frac{1}{\mu_{i}}+\frac{1}{\mu_{n}}$,
  the determinant of the metric is
\begin{align*}
&\det(g(D))=\left(\prod_{1\leq i<j\leq n}2m_{ij}\right)\varphi,\qquad\mbox{if}\ D\in\MR{n} \\
&\det(g(D))=\left(\prod_{1\leq i<j\leq n}4m_{ij}^{2}\right)\varphi,\qquad\mbox{if}\ D\in\MC{n},
\end{align*}
where $\varphi$ is a determinant of an $(n-1)\times(n-1)$ matrix
\begin{equation*}
\varphi=\det\begin{pmatrix}
\frac{1}{\mu_{1}}+\frac{1}{\mu_{n}} & \frac{1}{\mu_{n}} & \hdots & \frac{1}{\mu_{n}} \\
\frac{1}{\mu_{n}} & \frac{1}{\mu_{2}}+\frac{1}{\mu_{n}} & \hdots & \frac{1}{\mu_{n}} \\
\vdots & \vdots & \ddots & \vdots \\
\frac{1}{\mu_{n}} & \frac{1}{\mu_{n}} & \hdots & \frac{1}{\mu_{n-1}}+\frac{1}{\mu_{n}}
\end{pmatrix}=\frac{1}{\det(D)}.
\end{equation*}

\begin{theorem}
The volume of the real and complex state space endowed with a Riemannian metric which is generated by the
  operator monotone function $f$ is
\begin{align*}
&V(\MR{n},g_{f})=2^{\frac{n(n-1)}{4}}\hskip-0.5em\int\limits_{\MR{n}}\hskip-0.5em
  \frac{1}{\sqrt{\det(D)}}\left(\prod_{1\leq i<j\leq n}m(\mu_{i}(D),\mu_{j}(D))^{\frac{1}{2}}\right)
  \dint\la_{\dim(\MR{n})}(D)\\
&V(\MC{n},g_{f})=2^{\frac{n(n-1)}{2}}\hskip-0.5em\int\limits_{\MC{n}}\hskip-0.5em
  \frac{1}{\sqrt{\det(D)}}\left(\prod_{1\leq i<j\leq n}m(\mu_{i}(D),\mu_{j}(D))\right)
  \dint\la_{\dim(\MC{n})}(D).
\end{align*}
\end{theorem}

We can endow the Riemannian space with a pull-back metric too.
Consider the functions $h:\left\rbrack 0,1\right\lbrack\to\mathbb{R}$ with analytic continuation on a
  neighborhood of the $\left\rbrack 0,1\right\lbrack$ interval and suppose that $h'(x)\neq 0$ for every
  $x\in\left\rbrack 0,1\right\lbrack$.
We call such functions admissible functions.
The space $M_{n}$ will geometrically be considered a Riemannian space $(\mathbb{R}^{d},g_{E})$,
  where $d_{\mathbb{R}}=\frac{(n-1)(n+2)}{2}$ for real matrices and $d_{\mathbb{C}}=n^{2}-1$ for complex ones
  and $g_{E}$ is the canonical Riemannian metric on $M_{n}$.
That is, at every point $D\in\MN$ for every vectors $X,Y\in\MN$ in the tangent space at $D$ the metric is
\begin{equation*}
g_{E}(D)(X,Y)=\Tr XY.
\end{equation*}
For an admissible function $h:\left\rbrack 0,1\right\lbrack\to\mathbb{R}$ the pull back-geometry of the
  space $\MR{n}$ and $\MC{n}$ is the Riemannian geometry $g_{h}$ induced by the map
\begin{equation*}
\phi_{h,n}:\MN\to M_{n}\qquad D\mapsto h(D).
\end{equation*}
This Riemannian space will be denoted by $(\MN,g_{h})$.

For example if the functions $h$ are $p\root{p}\of{x}$ if $p\neq 0$ or $\log x$
  then we get the $\alpha$-geometries \cite{GibIso1,GibPis}.

If $D\in\MN$ is diagonal, i.e. $\displaystyle D=\sum_{i=1}^{n}\mu_{i}E_{ii}$, then the metric can be
  computed (see \cite{And3}) as
\begin{align*}
&\mbox{if} &&1\leq i<j\leq n, 1\leq k<l\leq n: \quad
  &&\left\lbrace\begin{array}{l} g(D)(H_{ij},H_{kl})=\delta_{ik}\delta_{jl}2M(\mu_{i},\mu_{j})^{2}\\
                               g(D)(F_{ij},F_{kl})=\delta_{ik}\delta_{jl}2M(\mu_{i},\mu_{j})^{2}\\
                               g(D)(H_{ij},F_{kl})=0,\end{array}\right.\\
&\mbox{if} &&1\leq i<j\leq n, 1\leq k\leq n: \quad &&g(D)(H_{ij},F_{kk})=g(D)(F_{ij},F_{kk})=0,\\
&\mbox{if} &&1\leq i\leq n,   1\leq k\leq n: \quad &&g(D)(F_{ii},F_{kk})=\delta_{ik}4M(\mu_{i},\mu_{i})^{2},
\end{align*}
where
\begin{equation*}
M(\mu_{i},\mu_{j})=\left\lbrace\begin{array}{ll}
  \dfrac{h(\mu_{i})-h(\mu_{j})}{\mu_{i}-\mu_{j}} &\quad\mbox{if}\ \mu_{i}\neq\mu_{j}\\
  h'(\mu_{i}) &\quad\mbox{if}\ \mu_{i}=\mu_{j}.\end{array}\right.
\end{equation*}
Using the previous considerations about the volume of the state space we have the following Theorem.

\begin{theorem}
The volume of the real and complex state space endowed with a pull-back metric $g_{h}$ is
\begin{align*}
&V(\MR{n})=2^{\frac{n(n-1)}{4}}\hskip-0.5em\int\limits_{\MR{n}}\hskip-0.5em
  \sqrt{\sum_{i=1}^{n}\prod_{\btop{j=1}{j\neq i}}^{n}h'(\mu_{j})}
  \left(\prod_{1\leq i<j\leq n}\hskip-1em M(\mu_{i}(D),\mu_{j}(D))\right)
  \dint\la_{\dim(\MR{n})}(D)\\
&V(\MC{n})=2^{\frac{n(n-1)}{2}}\hskip-0.5em\int\limits_{\MC{n}}\hskip-0.5em
  \sqrt{\sum_{i=1}^{n}\prod_{\btop{j=1}{j\neq i}}^{n}h'(\mu_{j})}
  \left(\prod_{1\leq i<j\leq n}\hskip-1em M(\mu_{i}(D),\mu_{j}(D))\right)^{2}
  \hskip-0.5em\dint\la_{\dim(\MC{n})}(D).
\end{align*}
\end{theorem}

\section{Volume of the state space of qubits}

In the space of qubits we choose the Stokes parametrization, i.e. we write a state $D\in\Mt$ in the form
\begin{equation*}
D=\frac{1}{2}\begin{pmatrix} 1+x & y+\ci z \\
                             y+\ci z & 1-x \end{pmatrix}.
\end{equation*}
Using these coordinates the spaces $\MC{2}$ and $\MR{2}$ can be identified with the
  unit ball in the Euclidean spaces $\mathbb{R}^{3}$ and $\mathbb{R}^{2}$.
The metric $g_{f}$ which is generated by an operator monotone function in this coordinate system is
\begin{equation*}
g_{f}(x,y,z)=\begin{pmatrix} \frac{1}{4\la_{1}\la_{2}} & 0 & 0\\
                             0 & \frac{m(\la_{1},\la_{2})}{2} & 0\\
                             0 & 0 & \frac{m(\la_{1},\la_{2})}{2} \end{pmatrix}\qquad
g_{f}(x,y)=\begin{pmatrix} \frac{1}{4\la_{1}\la_{2}} & 0 \\
                             0 & \frac{m(\la_{1},\la_{2})}{2}\end{pmatrix}.
\end{equation*}
The volume is an integral on the unit ball, which is in spherical and polar coordinates
\begin{align*}
&V(\MC{2})=4\pi\int_{0}^{1}\frac{r^{2}}{\sqrt{1-r^{2}}(1+r)f\left(\frac{1-r}{1+r} \right) }\dint r,\\
&V(\MR{2})=2\pi\int_{0}^{1}\frac{r}{\sqrt{1-r}(1+r)\sqrt{f\left(\frac{1-r}{1+r} \right)}}\dint r,
\end{align*}
respectively.

\begin{corollary}
The volume of the space $(\Mt,g_{f})$ where the metric $g_{f}$ is generated by an operator monotone
  function $f$ is
\begin{align*}
&V(\MC{2})=2\pi\int_{0}^{1}\left(\frac{1-t}{1+t}\right)^{2}\frac{1}{\sqrt{t}f(t)} \dint t\\
&V(\MR{2})=\sqrt{2}\pi\int_{0}^{1}\frac{1-t}{1+t}\frac{1}{\sqrt{t+t^{2}}\sqrt{f(t)}}  \dint t.
\end{align*}
\end{corollary}

Here are some operator monotone functions which generate monotone metrics from \cite{Pet1,Pet2,And1} and
  the corresponding volumes.
\begin{align*}
&f(x): && V(\MC{2}): && V(\MR{2}):\\
&\frac{1+x}{2} && \pi^{2} && 2\pi\\
&\frac{2x}{1+x} && \infty  && \infty\\
&\frac{x-1}{\log x} && 2\pi^{2}&& \sim 8.298\\
&\sqrt{x} && \infty && 4\pi\\
&\frac{1}{4}(\sqrt{x}+1)^{2} && 4\pi(\pi-2) && 4\pi(2-\sqrt{2})\\
&\frac{2\sqrt{x}(x-1)}{(1+x)\log x} && \infty && \sim 19.986\\
&\frac{2(x-1)^{2}}{(1+x)(\log x)^{2}} &&\frac{\pi^{4}}{2} && \sim 11.51\\
&\frac{x}{2}\left(\frac{1}{\alpha x+1-\alpha}+\frac{1}{(1-\alpha)x+\alpha} \right) && \infty && \infty\\
&\frac{2}{x+1}(\beta x+1-\beta)((1-\beta)x+\beta)
  && \pi^{2}\frac{1-2\sqrt{\beta-\beta^{2}}}{(1-2\beta)^{2}\sqrt{\beta-\beta^{2}}} && ?<\infty\\
&\frac{2x^{\gamma+\frac{1}{2}}}{1+x^{2\gamma}} && \infty && ?<\infty
\end{align*}
The parameters lie in the interval $\alpha\in\left]0,\frac{1}{2}\right]$,
  $\beta\in\left]0,\frac{1}{2}\right[$ and $\gamma\in\left[0,\frac{1}{2}\right]$.
We have some open questions about this list.
For every function in this list the complex state space has a greater volume;
  a natural question is if it is necessary?
It seems that if for a function $f$ the volume $V(\MC{2})$ is finite then for the transpose function
  $f^{\perp}(x)=\frac{x}{f(x)}$ the volume is infinity, except for the function
  $f(x)=\sqrt{x}$, in this case $f=f^{\perp}$.
Is it true in general?
For some functions the difference between the volumes is infinity.
What can be the statistical meaning of this phenomenon?

The origin of the infinite volume of the space $\MC{2}$ can be understood partially by the help of a
  representation theorem for operator monotone functions.
This representation Theorem is due to L\"owner \cite{Low}, but we use a modified version from \cite{GibIso3}.

\begin{theorem} The map $\mu\mapsto f$, defined by
\begin{equation*}
f(x)=\int_{0}^{1}\frac{x}{(1-t)x+t}\ d\mu(t),\qquad\mbox{for}\quad
x>0,\label{eq:f repr}
\end{equation*}
  establishes a bijection between the class of positive Radon measures on
  $\lbrack 0,1 \rbrack$ and the class of operator monotone functions.
The function $f$ fulfills the condition $f(x)=xf(x^{-1})$ for every positive $x$
  if and only if for every $s\in\left[0,1\right]$ the equality
  $\mu(\left[0,s \right])=\mu(\left[1-s,1 \right])$ holds.
\end{theorem}

If $f$ is an operator monotone function then it's transpose $f^{\perp}$ is monotone too \cite{Bha}.
That is, $1/f(x)$ can also be written in the form
\[\int_{0}^{1}\frac{1}{(1-t)x+t}\dint \mu(t),\]
  where $\mu$ is a probability measure on $\left[0,1 \right]$ with the
  symmetric property $\mu(\left[0,s \right])=\mu(\left[1-s,1 \right])$.
Substituting this representation of $f$ into the volume formula for $\MC{2}$ we have that
  if $\mu$ is the corresponding symmetric measure for the function $f^{\perp}$, then
  the volume of the manifold $(\MC{2},g_{f})$ is
\begin{equation*}
V=\int_{0}^{1}\frac{2}{2z-1}-\frac{\pi}{(2z-1)^{2}}
 +\frac{\arccos(2z-1)}{(2z-1)^2\sqrt{z-z^{2}}} \dint\mu(z).
\end{equation*}
The integrand is continuous, monotonously decreasing and has a series expansion
\begin{equation*}
\pi\frac{1}{\sqrt{z}}-(4+\pi)+\frac{9\pi}{2}\sqrt{z}-4\left(\frac{10}{3}+\pi \right)z+\dots
\end{equation*}
  near the origin.
It's integral with respect to a symmetric probability measure is infinity if and only if
\begin{equation*}
\int_{0}^{1}\frac{1}{\sqrt{z}}\dint\mu(z)=\infty
\end{equation*}
  holds.
So the volume of the complex state space of qubits is infinity if the metric is generated by
  a symmetric probability measure which is concentrated in some sense at the ends of the
  interval $\left[ 0,1\right]$.

If we consider the space of qubits with a pull-back metric, then using the above mentioned
  techniques we have the following corollary.

\begin{corollary}
For an admissible function $f$ let us consider the real and complex space $\Mt$ with the
  pull-back metric.
The volume of this space is the following.
\begin{align*}
&V(\MR{2})=\frac{\pi}{\sqrt{2}}\int_{0}^{1}\hskip-0.5em
  \sqrt{f'\left(\frac{1+r}{2}\right)^{2}+f'\left(\frac{1-r}{2}\right)^{2}}\hskip-0.5em
  \left(f\left(\frac{1+r}{2}\right)-f\left(\frac{1-r}{2}\right)\right)
  \dint r\\
&V(\MC{2})=\pi\int_{0}^{1}\hskip-0.5em
  \sqrt{f'\left(\frac{1+r}{2}\right)^{2}+f'\left(\frac{1-r}{2}\right)^{2}}\hskip-0.5em
  \left(f\left(\frac{1+r}{2}\right)-f\left(\frac{1-r}{2}\right)\right)^{2}
  \dint r
\end{align*}
\end{corollary}

\section*{Acknowledgements}
This work was supported by Hungarian Scientific Research Fund (OTKA)
  contract T046599, TS049835,
  and EU Network ''QP-Applications'' contract number HPRN--CT--2002--00729.

\end{document}